\documentclass[conference]{IEEEtran}
\IEEEoverridecommandlockouts

\IEEEoverridecommandlockouts
\usepackage{cite}
\usepackage{tikz}
\usepackage{amsmath,amssymb,amsfonts,amsthm}
\usepackage{authblk}
\usepackage{breqn}
\usepackage{algorithmic}
\usepackage{graphicx}
\usepackage{textcomp}
\usepackage{xcolor}
\usepackage{import}
\usepackage{multicol}
\usepackage{subfigure}
\usepackage{caption}
\usepackage{nicefrac}
\def\BibTeX{{\rm B\kern-.05em{\sc i\kern-.025em b}\kern-.08em
		T\kern-.1667em\lower.7ex\hbox{E}\kern-.125emX}}
\begin{document}
	
	\title{The Wideband Analysis of the Impact of I/Q Imbalance on THz Communication\\
		\thanks{This work was funded by the German Federal Ministry of Education and Research (BMBF) in the course of the 6GEM research hub under grant number 16KISK037.}
	}

\author[1]{Dogus Can Sevdiren}
\author[1]{Aydin Sezgin}
\author[2]{Mohammad Soleymani}

\affil[1]{\ Institute for Digital Communication Systems, Ruhr-Universit\"at Bochum, Germany}
\affil[2]{ Signal and System Theory Group, Universit\"at Paderborn, Germany}
\affil[ ]{ Emails: \{dogus.sevdiren, aydin.sezgin\}@rub.de,  mohammad.soleymani@uni-paderborn.de}

	\maketitle
\begin{abstract}
The terahertz (THz) band is a promising solution
to the increasing data traffic demands of future wireless networks. However, developing transceivers for THz communication is a complex and toilsome task due to the difficulty in designing devices that operate at this frequency and the impact of hardware impairments on performance. This paper investigates the impact of radio frequency (RF) impairment, in-phase/quadrature imbalance (IQI). To this end, we express an IQI model for the THz-specific array-of-subarrays (AoSA) architecture considering the unique features of THz communication; vast bandwidth, severe power drawdown, and pencil-like beams. We further model the impact of IQI in the power limited regime in order to investigate the power and ultra-wideband trade-off. To achieve this, we express the spectral efficiency in terms of wideband slope and bit energy to noise ratio which are the two important information theoretic metrics that reveals the performance of the ultra-wideband systems as in THz communication. Our results show that THz systems with IQI have a strict limit in achievable rate although they provide immense spectrum. We also demonstrate with our simulation results that compared to low frequencies, IQI is a more serious concern in THz links.
\end{abstract}


\section{Introduction}
\begin{figure*}[th!]
	\scalebox{0.40}
	\normalsize
	\setcounter{equation}{2}
	\begin{equation}
		\label{IQI_Overall}
		\begin{aligned}
			\mathbf{y}[k] &= \mathbf{W}_R[k] \left(\mathbf{K}_1[k]\mathbf{F}_R^H\mathbf{H}\left[k\right]\mathbf{F}_T\mathbf{G}_1[k] + \mathbf{K}_2[k]\mathbf{F}_R^{T}\mathbf{H}^{*}\left[-k\right]\mathbf{F}^{*}_T\mathbf{G}_2[k]\right)\mathbf{W}_T\left[k\right]\mathbf{s}\left[k\right] \\
			&+\mathbf{W}_R\left[k\right] \left(\mathbf{K}_1[k]\mathbf{F}_R^H\mathbf{H}\left[k\right]\mathbf{F}_T\mathbf{G}_2^{*}[k] + \mathbf{K}_2[k]\mathbf{F}_R^{T}\mathbf{H}^{*}\left[-k\right]\mathbf{F}^{*}_T\mathbf{G}_1^{*}[k]\right)\mathbf{W}_T^{*}\left[-k\right]\mathbf{s}^{*}\left[-k\right]\\
			&+\mathbf{W}_R\left[k\right]\left(\mathbf{K}_1[k]\mathbf{z}[k]+\mathbf{K}_2[k]\mathbf{z}^{*}[-k]\right)
		\end{aligned}
	\end{equation}
	\setcounter{equation}{0}
	\hrulefill
\end{figure*}
The growing reliance on online services has raised expectations for wireless communication systems, driven by a sharp increase in data traffic. Projected wireless network demand is set to exceed one terabit per second, with user-experienced data rates expected to be 10 to 100 times higher. To meet this demand, leveraging the unpolluted terahertz (THz) spectrum is crucial. However, despite significant efforts, developing efficient THz transceivers remains a challenging task.
\par
One significant challenge is the insufficient strength of the signal emitted by antennas operating in the THz band, reported in \cite{blackledge2021fractal}. Given this power gap, the ultra-wide bandwidth and the decreasing signal strength with distance, the power per unit bandwidth vanishes for THz systems. This case particularly leads THz systems to suffer from lack of power and operate in low signal-to-noise-ratio (SNR) regime. Another important challenge is, obtaining linear hardware response for ultra-wide bandwidth, which in turn introduces severe imperfections in the radio frequency (RF) domain, leads to degradation. One major impairment reported for the THz transceivers with is the in-phase/quadrature imbalance (IQI) \cite{SpatIQI}. More precisely, in \cite{300GMMIC} and \cite{VasquezIQrec}  attenuation levels around 20 decibel (dB) between in-phase (I) and quadrature (Q)  were reported for THz band. For comparison with the conventional frequencies, the Long-Term Evaluation (LTE) standard set forth by 3GPP, stipulated that user equipment must achieve a minimum attenuation of 25 or 28 dB \cite{KorpiValkama}.
\par
 Some aspects of IQI has been addressed in the state of the art already. For instance, in \cite{JavecIQI} and in \cite{MoK} authors discussed IQI for multiple-input-multiple-output (MIMO) systems. In \cite{MoImp}, the authors designed improper signals in order to mitigate the impact of IQI. In \cite{SpatIQI}, the THz spatial modulation in the presence of IQI was investigated. In \cite{PrecompIQI}, authors proposed a pre-compensation scheme to compensate for nonlinearity and IQI effects. In \cite{ChannelEstIQI}, authors developed a channel estimation and equalization method for THz receivers in the presence of IQI. 
\par
This study explores the impact of IQI on a multi-user (MU)-MIMO THz communication system using orthogonal frequency division multiplexing (OFDM) and hybrid beamforming. We employ the wideband slope \(\mathcal{S}_0\) and bit energy-to-noise ratio \(\nicefrac{E_b}{N_0}\) as information-theoretic metrics to assess spectral efficiency (SE) in the low SNR regime, highlighting the challenges posed by the power gap. The paper also compares IQI's effects with inter-user interference (IUI), addressing a gap in existing literature by focusing on ultra-wideband MU-MIMO systems at low power levels for THz communication. This research emphasizes the significant impact of IQI on system performance and the need for further investigation into optimizing power consumption and system efficiency under such conditions.
 
\section{IQI in Multicarrier Systems}
\begin{figure}[!htb]
\centering
\begin{minipage}[c]{0.45\textwidth}
	\centering
	\begin{tikzpicture}
		\draw[ultra thick, 	->,blue] (0.5,0) -- (0.5,1) node[anchor=south]{$-K$};
		\filldraw [blue] (0.75 ,0.5) circle (1pt);
		\filldraw [blue] (1,0.5) circle (1pt);
		\filldraw [blue] (1.25,0.5) circle (1pt);
		\draw[ultra thick, 	->, blue] (1.5,0) -- (1.5,1) node[anchor=south]{$-k$};;
		\filldraw [blue] (2 ,0.5) circle (1pt);
		\filldraw [blue] (2.25,0.5) circle (1pt);
		\filldraw [blue] (2.5,0.5) circle (1pt);
		\draw[ultra thick, 	->, blue] (3,0) -- (3,1)node[anchor=south]{$-2$};;
		\draw[ultra thick, 	->, blue] (3.60,0) -- (3.60,1)node[anchor=south]{$-1$};;
		\draw[ultra thick, 	->] (4,0) -- (4,2) node[anchor=south]{$f_c $};
		\draw[ultra thick, 	->,red] (7.5,0) -- (7.5,1) node[anchor=south]{$K$};
		\filldraw [red] (7.25 ,0.5) circle (1pt);
		\filldraw [red] (7,0.5) circle (1pt);
		\filldraw [red] (6.75,0.5) circle (1pt);
		\draw[ultra thick, 	->, red] (6.5,0) -- (6.5,1) node[anchor=south]{$k$};;
		\filldraw [red] (6 ,0.5) circle (1pt);
		\filldraw [red] (5.75,0.5) circle (1pt);
		\filldraw [red] (5.5,0.5) circle (1pt);
		\draw[ultra thick, 	->, red] (5,0) -- (5,1)node[anchor=south]{$2$};;
		\draw[ultra thick, 	->, red] (4.40,0) -- (4.40,1)node[anchor=south]{$1$};;
		\draw[ultra thick, <->] (0,0) -- (8,0) node[anchor=west]{$f$};	
	\end{tikzpicture} 
	\caption{Subcarrier indexing on passband}
	\label{Subcarriers}	
 \end{minipage}
\end{figure}
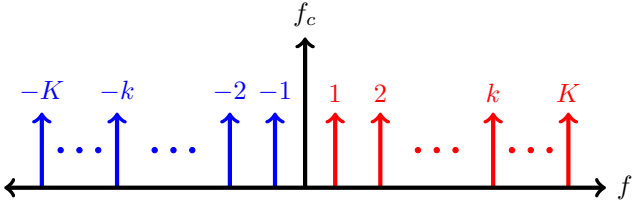
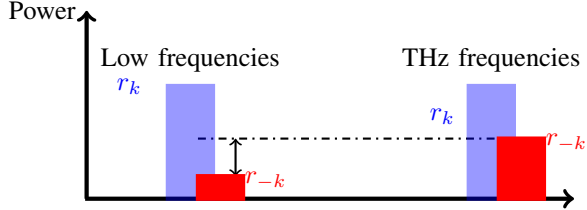
\begin{figure}[!htb]
	\centering
	\begin{tikzpicture}
		\draw[ultra thick, ->] (-.5,0) -- (6,0) node[anchor=west]{};	
  		\draw[ultra thick, ->] (-.48,0) -- (-.48,2.5) node[anchor=east]{Power};
  		\draw[ thick, dash dot] (1, .8) -- (5, .8) node[anchor=west ]{};	
  		\draw[ thick, <->] (1.5, .3) -- (1.5, .8) node[anchor=west]{};	
		\draw[ultra thick, blue, fill, opacity = .4] (0.6,0) rectangle (1.2,1.5);
		\draw[ultra thick, blue,  draw opacity=0] (0.1,0.7) rectangle (0.4,1.5) node[anchor=east] {$r_{k}$};
		\draw[ultra thick, red, fill, fill opacity=1] (1,0) rectangle (1.6,.3);
		\draw[ultra thick, red, draw opacity=0] (1.7,0) rectangle (2.1,.5) node[pos=.5] {$r_{-k}$};
		\draw[black, draw opacity=0] (0.6,1.7) rectangle (1.2,2) node[pos=.5] {Low frequencies};	
		\draw[ultra thick, blue, fill, opacity = .4] (4.6,0) rectangle (5.2,1.5);
		\draw[ultra thick, blue,  draw opacity=0] (4.1,0.7) rectangle (4.4,1.5) node[pos=.5] {$r_{k}$};
		\draw[ultra thick, red, fill, fill opacity=1] (5,0) rectangle (5.6,.8);
		\draw[ultra thick, red, draw opacity=0] (5.7,0) rectangle (6.1,1.5) node[pos=.5] {$r_{-k}$};
		\draw[black, draw opacity=0] (4.6,1.7) rectangle (5.2,2) node[pos=.5] {THz frequencies};
	\end{tikzpicture} 
	\caption{Comparison of IQI between THz and low frequencies}
	\label{contaminated}	
 \vspace{5mm}
\end{figure}
In an OFDM system, the mismatches in the I and Q branches of the local oscillator (LO) signal during up-/down conversion can cause interference among subcarriers. For an OFDM signal with $2K$ subcarriers, $ k \in \{-K, ..., -1, 1, ..., K\}$, where subcarriers are symmetric around the carrier frequency \(f_c\) (Fig. \ref{Subcarriers}), IQI results in interference from the symmetric image band during downconversion at the receiver \cite{schenk2006estimation} (Fig. \ref{contaminated}). Notably, the power of the interfering image band is higher at THz frequencies than at lower frequencies, and the same issue occurs during upconversion at the transmitter
\section{System Model}
Consider a MU-MIMO OFDM broadcast channel, where ttransceivers have the structure of array-of-subarrays (AoSA) \cite{AoSA}, with a transmitter equiped with $N$ SAs and $M$ single SA users. An SA is composed of $Q$ antenna elements (AE)s where each AE has its own phase shifter and it is linked to a dedicated RF chain. We can express received passband signal $\mathbf{r}\in\mathbb{C}^{MQ\times 1}$ as
\begin{equation}
    \mathbf{r}[k] = \mathbf{H}[k]\mathbf{x}[k]+\mathbf{z}[k],
\end{equation}
where $k$ denotes the subcarrier index out of the $2K$ subcarriers depicted in the Fig. \ref{Subcarriers}. 
Taking into account the presence of IQI in both the transmitter and receiver, we can represent the transmitted passband signal and the received baseband signal as,
\begin{subequations}
	\setlength{\arraycolsep}{0.0em}
	\begin{eqnarray}  
		\hspace{-5mm}\mathbf{x}[k] &=& \mathbf{F}_T(\mathbf{G}_1[k]\mathbf{W}_T[k]\mathbf{s}[k] + \mathbf{G}_2^*[k]\mathbf{W}_T^*[-k]\mathbf{s}^*[-k]), \\
		\hspace{-5mm}\mathbf{y}[k] &=& \mathbf{W}_R[k](\mathbf{K}_1[k]\mathbf{F}_R^H \mathbf{r}[k] + \mathbf{K}_2[k]\mathbf{F}_R^T\mathbf{r}^*[-k]).
	\end{eqnarray}
\end{subequations}
Here, $\mathbf{W}_R\in\mathbb{C}^{M\times M}$ and $\mathbf{F}_R\in\mathbb{C}^{MQ\times M}$, respectively, represent the digital and analog combiners, at the receiver side $\mathbf{W}_T\in\mathbb{C}^{N\times N}$ and $\mathbf{F}_T\in\mathbb{C}^{NQ\times N}$, respectively, indicate the digital and analog beamformers, at the transmitter side. $\mathbf{H}\in\mathbb{C}^{MQ\times NQ}$  is the THz channel matrix, and $\mathbf{z}\in\mathbb{C}^{MQ\times 1} $ is the circularly symmetric complex Gaussian noise vector. Moreover, the transmit symbol vector is denoted by. The overall system model is given in (\ref{IQI_Overall}) at the top of the page.
 \par 
 The matrices, $\mathbf{G_1}$, $\mathbf{G_2}$, $\mathbf{K_1}$, and $\mathbf{K_2}$ capture the amplitude and rotational imbalance associated with the transmitter and receiver, respectively, and are given by \cite{schenk2006estimation}
\setcounter{equation}{3}
\begin{subequations}
	\setlength{\arraycolsep}{0.0em}
	\begin{eqnarray}
		\mathbf{G_1}&=& \dfrac{1}{2}(\mathbf{I} + \mathbf{G}_Te^{j\mathbf{\Phi}_T}), \\
		\mathbf{G}_2 &=& \mathbf{I} - \mathbf{G}_1^*,\\
		\mathbf{K_1} &=& \dfrac{1}{2}(\mathbf{I} + \mathbf{G}_R e^{-j\mathbf{\Phi}_R}), \\ 
		\mathbf{K}_2 &=& \mathbf{I} - \mathbf{K}_1^*,
	\end{eqnarray}
\end{subequations}
where $\mathbf{I}$ is the identity matrix. The matrices $\mathbf{G}_T$ and $\mathbf{G}_R$   
reflect the amplitude errors, $\mathbf{\Phi}_T$ and $\mathbf{\Phi}_R$ reflect the phase errors at the transmitter and receiver, respectively. The error matrices are diagonal and given as \cite{JavecIQI}
\begin{subequations}
	\setlength{\arraycolsep}{0.0em}
	\begin{eqnarray}
		\mathbf{G}_x &= \mbox{diag}\{g_{x,1} g_{x,2}, ..., g_{x,N_x}\}, \\
		\mathbf{\Phi}_x &= \mbox{diag}\{\Phi_{x,1} \Phi_{x,2}, ..., \Phi_{x,N_x}\},
	\end{eqnarray}
\end{subequations}
where $x$ can be $T$ or $R$, denoting the transmitter and receiver sides, respectively. A system with perfect IQ response is represented by the values $g_{x,i}=1$ and $\Phi_{x,i} = 0$.
\subsection{Terahertz Channels}
THz channels have been widely studied in the literature \cite{OverviewTHz}. Signal propagation in the THz band is mainly dominated by a few reflection paths, with accompanying significant attenuation. As a result, THz channels are primarily influenced and modeled by the line-of-sight (LOS) path, and the channel between the \(n\)-th transmit SA and the \(m\)-th user can be written as\cite{OverviewTHz} 
\begin{equation}
	\label{channel_AoSA}
	\mathbf{H}_{mn}(f,\Delta) = \alpha_{mn}(f,\Delta)\mathbf{a}_{r}(\phi_{mn}^{a}, \theta_{mn}^{a})\mathbf{a}_{t}^{H}(\phi_{mn}^{d}, \theta_{mn}^{d})
\end{equation}
where $\alpha_{mn}$ denotes the path loss, $\mathbf{a}_r$ and $\mathbf{a}_t$ represent the receive and transmit antenna steering vectors, respectively. $\phi_a$, $\theta_a$ and $\phi_d$, $\theta_d$ denote the angles of arrival (AoA) and the angles of departure (AoD), respectively where $\phi$ corresponds to azimuth and $\theta$ corresponds to elevation. The path loss can be expressed as follows 
\begin{equation}
	\alpha_{mn} = G_TG_R\dfrac{c}{4\pi f\Delta}
\end{equation}
$f$ represents frequency, and $\Delta$ denotes the distance between the $n$-th transmit SA and the $m$-th user. Sequentially, $G_T$ and $G_R$ indicate the antenna gains of the transmitter and receiver.
\par
Furthermore, we assume that the communcation nodes have perfect spatioal knowledge.  In this case, the optimal analog beamformer between the \(m\)-th user and the \(n\)-th transmit SA is expressed using the array steering vector of the corresponding SAs 

\begin{subequations}
	\setlength{\arraycolsep}{0.0em}
	\begin{eqnarray}
		\mathbf{f}_{r}^{m} && =\mathbf{a}_{r}(\phi_{mn}^{a}, \theta_{mn}^{a}),  \\
		\mathbf{f}_{t}^{n} && =\mathbf{a}_{t}^*(\phi_{mn}^{d}, \theta_{mn}^{d}),
	\end{eqnarray}
\end{subequations}
where $\mathbf{f}_{r}^{m}$ is the $m$-th column of the receive analog combiner $\mathbf{F}_R$, and   $\mathbf{f}_{t}^{n}$ is the $n$-th column of analog beamformer $\mathbf{F}_T$. The array steering vector of a uniform rectangular planar array (URPA) is \cite{OverviewTHz}
	\begin{equation}
	               \mathbf{a}_0(\phi_0,\theta_0) = \dfrac{1}{\sqrt{Q}} \left[ e^{jk\Phi_{1,1}}, ...,e^{jk\Phi_{u,v}},..., e^{jk\Phi_{Q,Q}}\right],
	\end{equation}
where ${\Phi_{u,v}}$ is the phase shift of the AE ($u,v$) and given as
\begin{dmath}
                 {\Phi_{u,v}}(\phi_0,\theta_0) = s_x^{u,v} \cos\phi_0\sin\theta_0 + s_y^{u,v} \cos\phi_0\sin\theta_0  + s_z^{u,v} \cos\theta_0,
\end{dmath}
where $s_x$, $ s_y$, $ s_z$ denoting the location of the corresponding AE in the directions $x$, $y$ and $z$ of the 3D coordinate system, respectively. 
\par 
The THz channel between the transmit and receive SAs exhibits high correlation due to the dominant line-of-sight (LOS) path and the small size of the antennas, leading to a rank-deficient channel matrix. By exploiting this rank deficiency and utilizing optimal beamforming, the channel dimensions between the transmitter and receiver can be reduced to those of a MIMO channel. This is achieved by concatenating the analog domain components, i.e., the channel and the analog beamformers, as
\begin{equation}
	\label{concatenated}
	\mathbf{H}_c[k] =  \mathbf{F}_R^{H}\mathbf{H}[k]\mathbf{F}_T,
\end{equation}
\subsection{Rate Expression with IQI}
The signal-to-interference-plus-noise ratio (SINR) serves as the key metric for assessing the performance of a communication system in high SNR regime and IQI introduces additional interference, as shown in Fig. \ref{contaminated}. In (\ref{IQI_Overall}), one can notice that the received signal consists of three components, the signal of the desired sub-carrier $\mathbf{s}[k]$, inter-carrier interference (ICI) $\mathbf{s}[-k]$, and the noise $\mathbf{z}[k]$. Hence, we can split the signal model into two parts, by separating the channel of the desired signal and the channel of the ICI. The channel of the desired sub-carrier signal is
\begin{equation}
	\label{desiredchannel}
	\mathbf{H}_d[k] = \mathbf{K}_1\mathbf{H}_c[k]\mathbf{G}_1 + \mathbf{K}_2\mathbf{H}_c[-k]^*\mathbf{G}_2,
\end{equation}
and the channel of the ICI is
\begin{equation}
	\label{interferencechannel}
	\mathbf{H}_i[k] = \mathbf{K}_1\mathbf{H}_c[k]\mathbf{G}_2^* + \mathbf{K}_2\mathbf{H}_c[-k]^*\mathbf{G}_1^*.
\end{equation}
Finally, we can define the rate of the $k$-th subcarrier by
\begin{equation}
	\hspace{-3mm} \mbox{R}[k] = \text{log}_2\left|\mathbf{I} + \mathbf{C}^{-1}\mathbf{W}_R^H\mathbf{H}_d\mathbf{W}_T[k]\mathbf{W}_T^H[k]\mathbf{H}_d^H\mathbf{W}_R  \right|,
\end{equation}
where $s_[k]$ is  circularly symmetric complex gaussian with $s_[k] \sim \mathcal{N}(0,\mathbf{I})$. Here we treat interference as noise and the corresponding interference plus noise covariance matrix $\mathbf{C}$ is defined as
\begin{equation}
\label{CovInt}
	\mathbf{C} = \mathbf{W}_R\mathbf{H}_i\mathbf{W}_T[-k]^*\mathbf{W}_T[-k]^T\mathbf{H}_i^H\mathbf{W}_R + \bar{\mathbf{Z}} ,
\end{equation}
where $\bar{\mathbf{Z}}$ denotes the noise covariance matrix with IQI, and written as
\begin{equation}
    \bar{\mathbf{Z}} = \dfrac{\sigma^2}{2}\left(\mathbf{I}+\mathbf{G_TG_T^H}\right)
\end{equation}
with
\begin{equation}
	E\{\mathbf{z}^H[k]\mathbf{z}[k]\} = {\sigma}^2\mathbf{I},
\end{equation}
where $\mathbf{I}$ to be equal to identity matrix and $G_x$ is drawn from amplitude error matrix of (5) for the receiver. We can observe from the covariance matrix of interference give in (\ref{CovInt}) that, the source of the interference is the signal located at the image frequency. In order to mitigate the interference due to IQI, the subcarrier associated with the image frequency can be nulled which implies that no signal is being transmitted on this specific subcarrier.
\section{Wideband Analysis}
In ultra-wideband systems in the power limited regime, the power per unit bandwidth diminishes around zero and system operates in low SNR regime. In this regime, the key measure for the system performance is the energy-per-information bit developed by Verd\'{u} in \cite{Verdu}, where SE is formulated in terms of energy per bit normalized to background noise spectral level $\nicefrac{E_b}{N_{0}}$ as
\begin{equation}
	\text{SE}\left(\dfrac{E_b}{N_0}\right) \approx \mathcal{S}_0 \dfrac{\dfrac{E_b}{N_0}(\mbox{dB}) - \dfrac{E_b}{N_{0, \min}}(\mbox{dB})}{\mbox{3dB}}.
\end{equation}
Here, $\nicefrac{E_b}{N_{0,\min}}$ is the minimum achievable energy per bit, and $\mathcal{S}_0$ is the wideband slope. 
\subsection{Wideband Slope in the Absence of IQI}
We first derive the wideband slope  and minimum achievable energy per bit of THz communication link in the absence of IQI. In \cite{SmallBWCase}, the authors presented a model for interference channels in low-SNR regime. By exploiting this model, we can express the minimum achievable energy per bit for the system without IQI. We can write the system without IQI as
\begin{equation}
    \mathbf{y}[k] = \mathbf{W}_R[k]\mathbf{H}_c[k]\mathbf{W}_T[k]\mathbf{s}[k] + \mathbf{W}_R[k]\mathbf{z}[k], 
\end{equation}
where $\mathbf{H}_C$ is the concatenated channel given in (11). For this system we can express the SINR of the $k$-th subcarrier for $m$-th user by treating interference as noise and under the equal power constraint as
\begin{equation}
    \gamma_m [k] = \frac{|h_{c,mm}[k]|^2 P}{P\sum_{n\neq m}|h_{c,mn}[k]|^2 + \sigma^2},
\end{equation}
where $P$ is the power, $h_{c,mn}$ is the element in the $m$-th row and $n$-th column of $\mathbf{H}_c$. Under the unitary noise power assumption, see \cite{SmallBWCase}, we can express the minimum achievable energy per bit as 
\begin{equation}
	\dfrac{E_b}{N_{0,\min}} = \dfrac{N2K\mbox{log}_e2}{\sum\limits_{\substack{k=-K \\ k\neq 0}}^K\sum\limits_{m=1}^{M}\big\lvert{h}_{c,mm}[k]\big\rvert^2},
\end{equation}
and the wideband slope as 
\begin{equation}
	\hspace{-3mm}\resizebox{0.92\hsize}{!}{
		$ \mathcal{S}_0 = \dfrac{2\left(\sum\limits_{\substack{k=-K \\ k\neq 0}}^K\sum\limits_{m=1}^{M}\big|{h}_{c,mm}[k]\big|^2\right)^2}
		{\sum\limits_{\substack{k=-K \\ k\neq 0}}^K\sum\limits_{m=1}^M\left( \big|{h}_{c,mm}[k]\big|^4 + \sum\limits_{\substack{n=1 \\ i\neq m}}^N \big|{h}_{c,mm}[k]\big|^2 \big|{h}_{c,mn}[k]\big|^2 \right)}$.
	}
\end{equation}
\subsection{Wideband Slope in the Presence of IQI}
We now extend this model to account for the presence of IQI and resulting interference. The minimum achievable energy per bit is affected by the presence of IQI. We can express the SINR of the $k$-th subcarrier for $m$-th user by treating interference as noise and under the equal power constraint as
\begin{equation}
    \gamma_m [k] = \frac{|h_{d,mm}[k]|^2 P}{P\left(\sum_{n\neq m}|h_{d,mn}[k]|^2 + \sum_{n=1}^{n=N}|h_{i,mn}[k]|^2\right) + \sigma^2},
\end{equation}
Here subscript $d$ denotes the entries of the desired signal channel matrix $\mathbf{H}_d$ given in (\ref{desiredchannel}), and subscript $i$ denotes the entries of the interference signal channel matrix $\mathbf{H}_i$ given in (\ref{interferencechannel}). Similarly, $m$ denotes the row, and $n$ denotes the column corresponding to the user and transmit SAs, respectively. By comparing (23) with (20), we can observe that IQI superpose an additional interference and modifies the end-to-end analog properties of the system, hence the channel. Under the same assumptions, we can express the minimum achievable energy per bit as
 \begin{equation}
	\dfrac{E_b}{N_{0,\min}}\bigg|_{IQI}= \dfrac{N2K\mbox{log}_e2}{\sum\limits_{\substack{k=-K \\ k\neq0}}^K\sum\limits_{m=1}^{M}\big\lvert{h}_{d,mm}[k]\big\rvert^2},
\end{equation}
and the wideband slope as
\begin{equation}
	\mathcal{S}_0\bigg|_{IQI} = 2\dfrac{\left(\sum\limits_{\substack{k=-K \\ k\neq0}}^K\sum\limits_{m=1}^{M}\big|{h}_{d,mm}[k]\big|^2\right)^2}
	{\sum\limits_{\substack{k=-K \\ k\neq0}}^K\sum\limits_{m=1}^M\left( \big|{h}_{d,mm}[k]\big|^4 + \zeta_m \right)},
\end{equation}
with
\begin{equation} \label{eq1}
	\begin{split}
		\zeta_m & =   \sum\limits_{\substack{n=1 \\ n\neq m}}^N \big|{h}_{d,mm}[k]\big|^2 \big|{h}_{d,mn}[k]\big|^2 \;\;\\
		   &      +\sum\limits_{n=1}^N \big|{h}_{d,mm}[k]\big|^2 \big|{h}_{i,mn}[k]\big|^2,
	\end{split}
\end{equation}
 where $\zeta_m$ represents the additional interference terms due to IQI.
\section{Numerical Simulation}
In the simulations, we consider a system communicating at $300$ GHz. We employ a transmitter with three SAs; and three single SA users. The results can be extended to multi-SA user cases without loss of generality. The number of antenna elements on a single SA is 256, i.e., $Q = 16$, $ N = 3$, and $ M = 3$. We use a bandwidth of $10$ GHz. In each simulation, users are located at arbitrary locations at a certain distance.
 \subsection{Result of Impact of IQI on Wideband Slope}

\begin{figure}[h!]
        \vspace{-2mm}
	\centering
	\includegraphics[width={0.40\textwidth}]{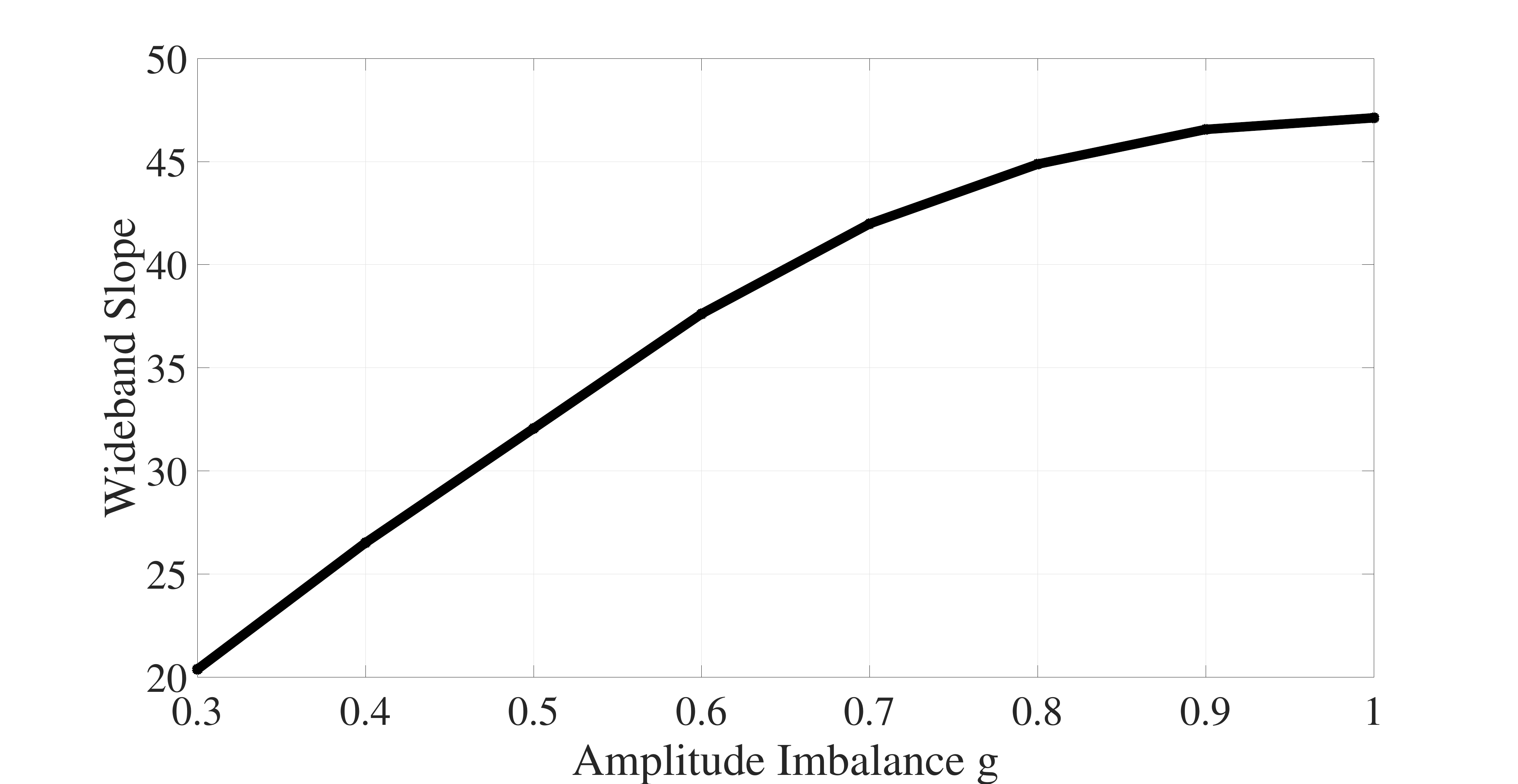}
	\caption{Wideband Slope vs. Amplitude Impairment}
	\label{fig:Slope}
    
	\centering
	\includegraphics[width={0.40\textwidth}]{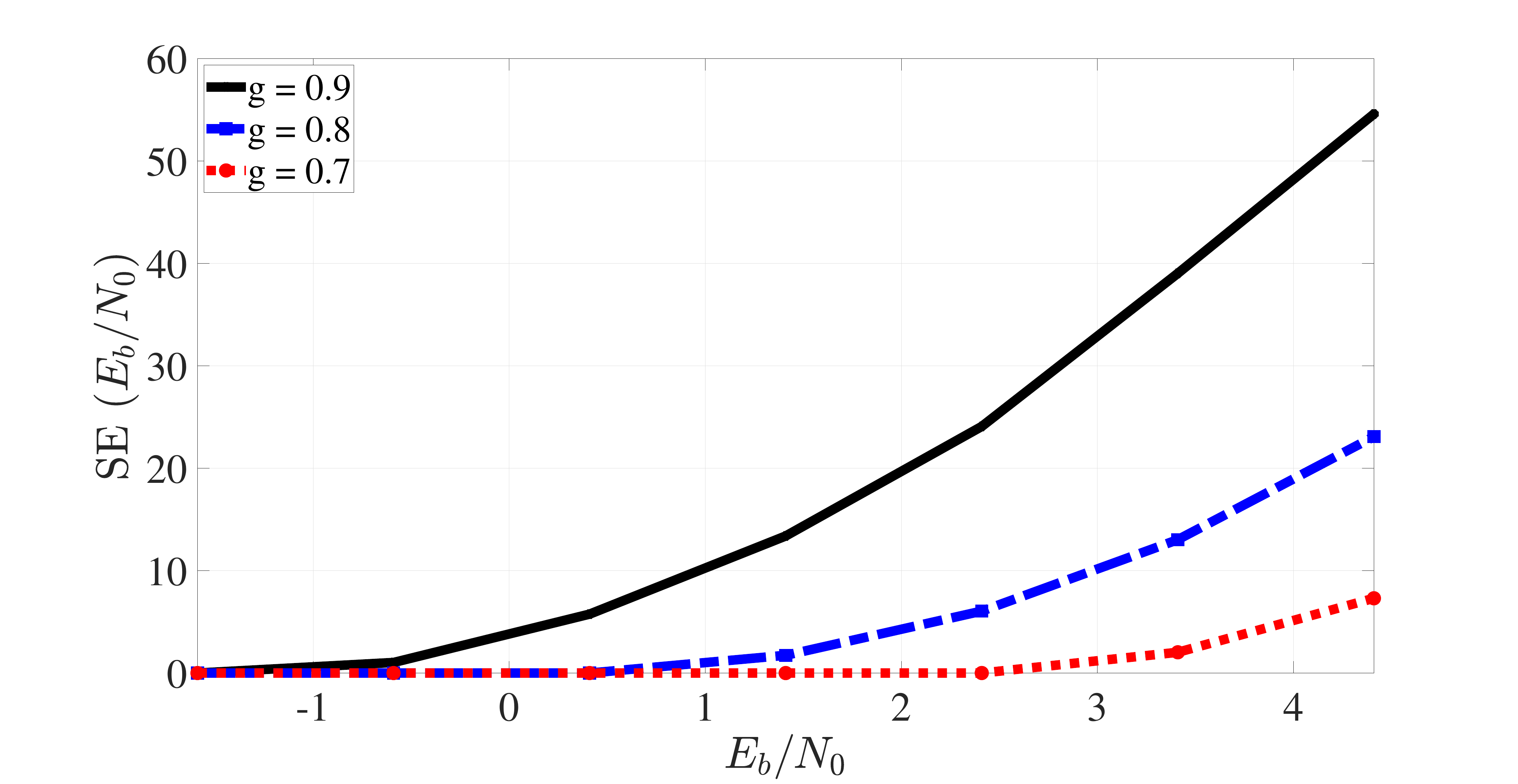}
	\caption{Spectral Efficiency vs. $E_b/N_0$}
	\label{fig:LowPowerCapacity}
            \vspace{-5mm}

\end{figure}

This subsection presents simulation results demonstrating the impact of IQI on a wideband, power-limited system. Specifically, Fig. \ref{fig:Slope} shows the achievable wideband slope with respect to IQI, measured in terms of amplitude imbalance. This can be interpreted as the system's benefit per unit of resource (e.g., power or bandwidth) in the presence of IQI. Lower values of $g$ indicate higher levels of IQI. The slope curve reveals that as IQI becomes more sever, the benefit the system get from increasing resources significantly diminishes.
\par
 Similarly, Fig. \ref{fig:LowPowerCapacity} illustrates the impact of the IQI on SE. In this figure, each curve corresponds to a different amplitude imbalance, i.e. solid black line shows $g = 0.9$, the blue dashed line shows $g=0.8$, and the red dotted line shows $g = 0.7$. As it can be observed, as the IQI becomes more severe, the bit energy required for the same perforamance is significantly increased.
\subsection{Analysis of IQI and Subcarrier Nulling}

\begin{figure}[t!]
	\centering
	\includegraphics[width=0.40\textwidth]{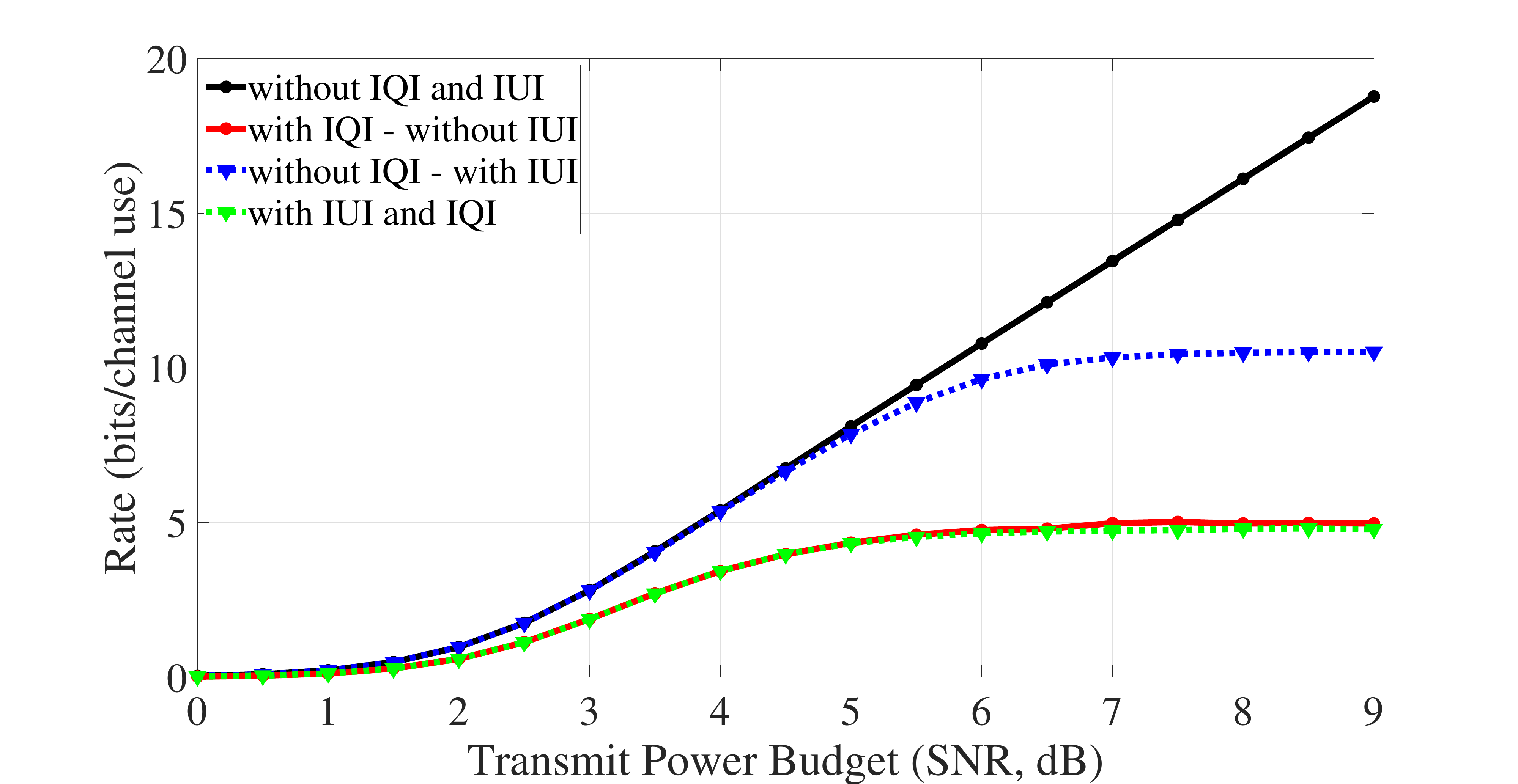}
	\caption{Impact of IQI on rate in THz band}
	\label{fig:THzIQISpec}
        \vspace{-.5mm}
	\centering
	\includegraphics[width=0.40\textwidth]{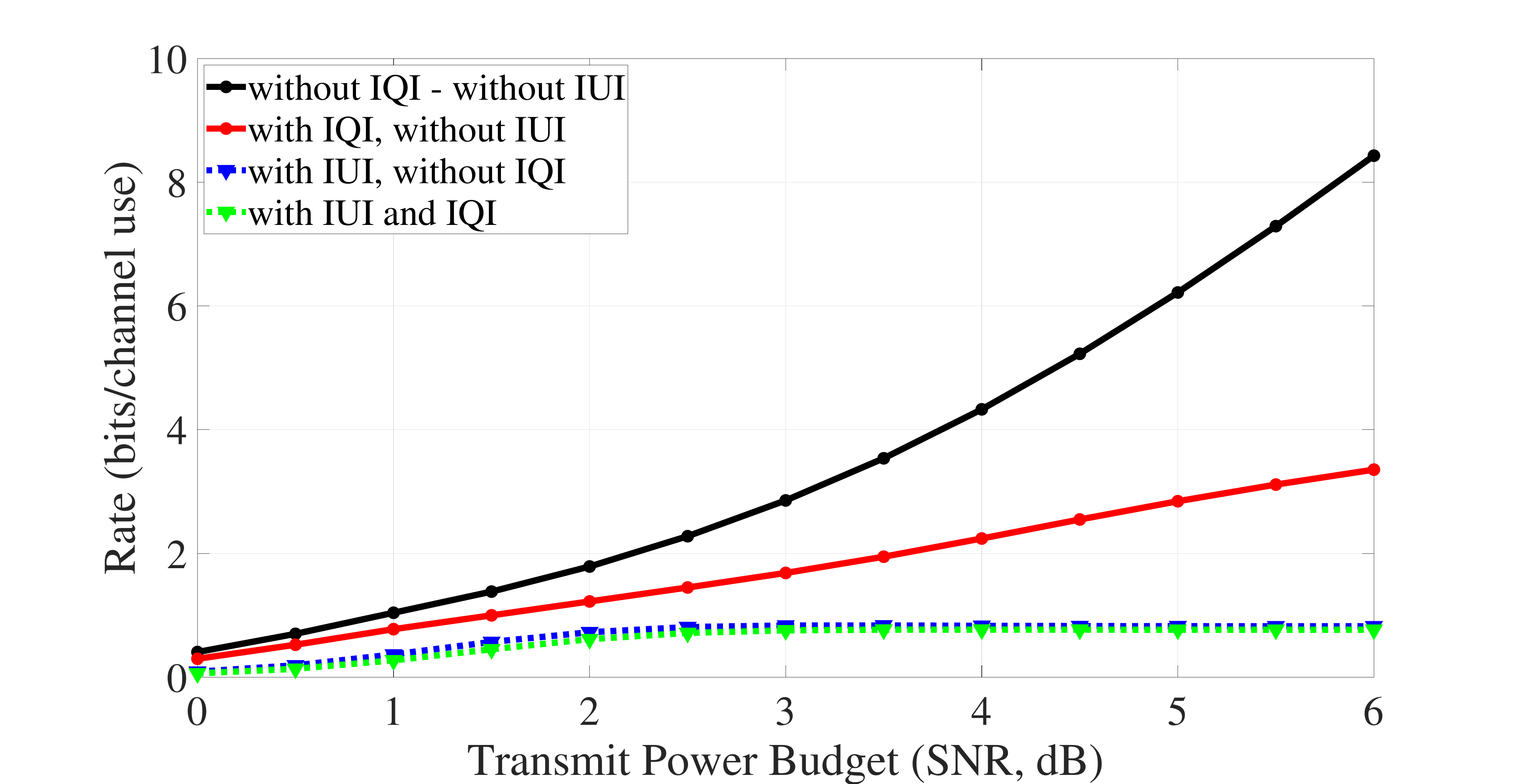}
	\caption{Impact of IQI on rate in Low Frequencies}
	\label{fig:ConvIQISpec}
    \vspace{-3mm}
\end{figure}
\par
This subsection examines the impact of IQI and subcarrier nulling on data rates in THz systems, using a phase imbalance of $5^{\circ}$ and an image rejection ratio of 30 dB, as specified in \cite{300GMMIC}. We analyze systems with and without IUI, using Rayleigh fading for conventional frequencies.

Fig. \ref{fig:THzIQISpec} shows the relationship between achievable data rates and power, given in SNR, in the presence of ICI from IQI and IUI. The solid black line indicates the rate without interference, showing a steady increase with power. The red line, with IQI, saturates as power increases due to growing IQI interference. The blue dashed line with IUI limits the achievable rate, while the red and green dashed lines (IQI only and joint IQI/IUI) overlap, suggesting IQI is more severe than IUI in THz systems. In contrast, Fig. \ref{fig:ConvIQISpec} shows that IUI is more problematic at lower frequencies, where systems can achieve higher rates despite IQI.

\begin{figure}[t!]
	\centering
	\includegraphics[width=0.40\textwidth]{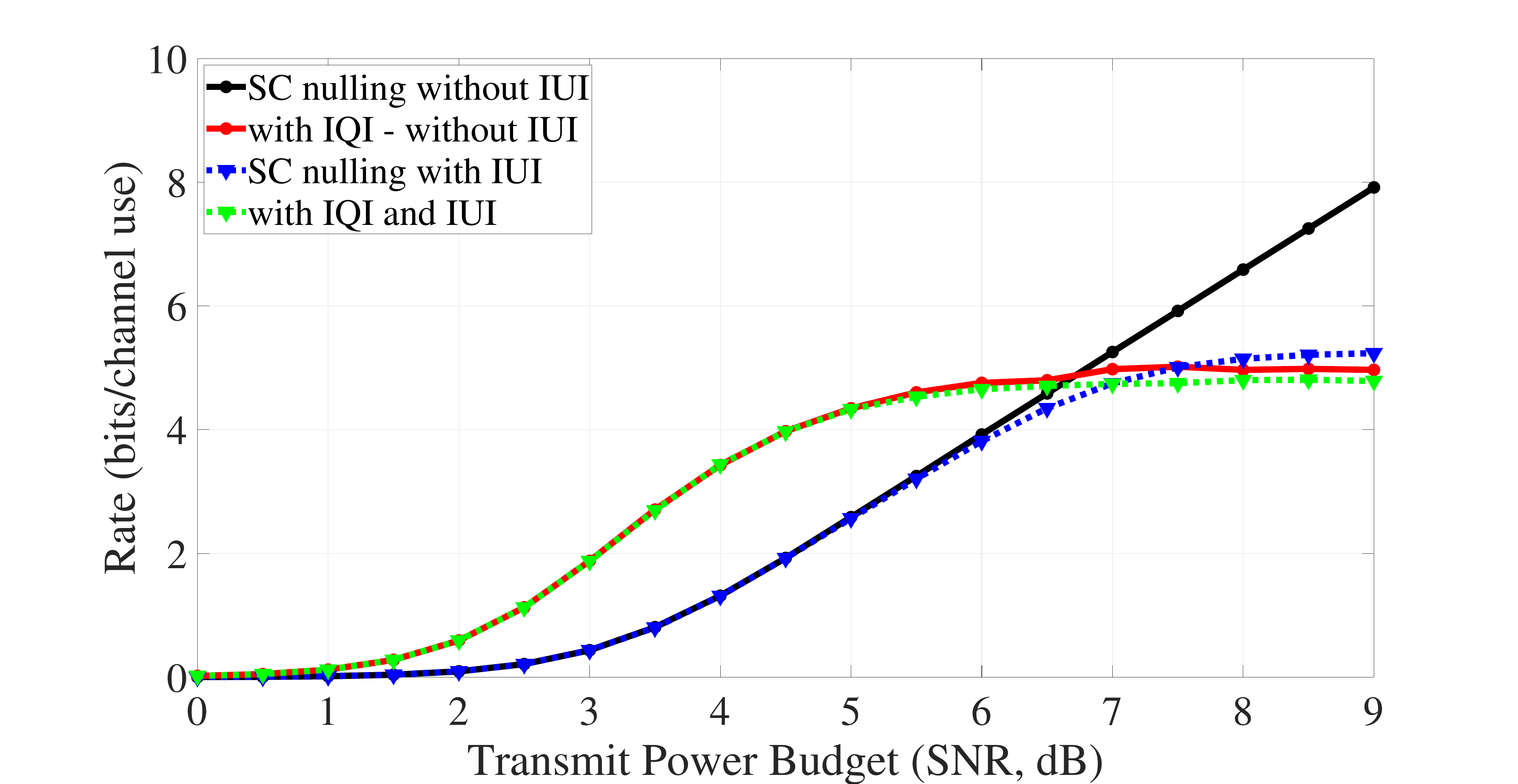}
	\caption{Subcarrier nulling THz band}
	\label{fig:THzSC}
            \vspace{-.5mm}
	\centering
	\includegraphics[width=0.40\textwidth]{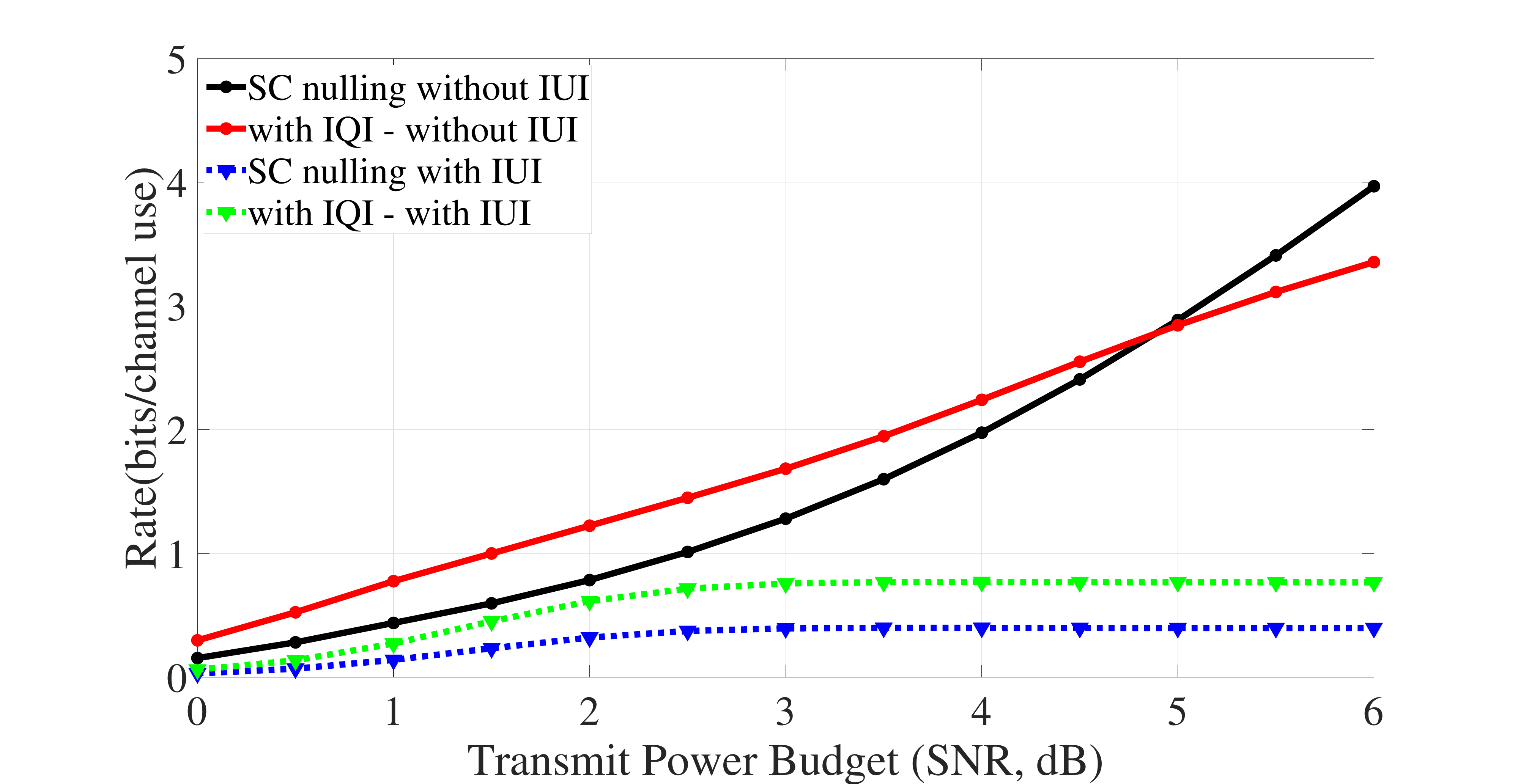}
	\caption{Subcarrier nulling in Low Frequencies}
	\label{fig:ConvSC}
        \vspace{-3mm}
\end{figure}

To mitigate IQI, we propose subcarrier nulling, where power is allocated to the subcarrier $k$ but not to the $-k$-th subcarrier, effectively halving the bandwidth to avoid intercarrier interference. Figs. \ref{fig:THzSC} and \ref{fig:ConvSC} show that subcarrier nulling consistently outperforms using the full bandwidth, particularly in THz systems where IUI is negligible.

An important finding is that, despite the abundant THz spectrum, data rates do not consistently increase with bandwidth when IQI is present. Using half the bandwidth can be more effective, highlighting IQI as a key obstacle to achieving THz communication’s potential.
\vspace{-2mm}
\section{Conclusion}
This research examines IQI in THz communication systems, highlighting its significant challenge and the need for effective mitigation strategies. Our analysis of power-limited wideband systems reveals the negative impact of IQI on spectral efficiency, emphasizing the need for further research to develop mitigation techniques and fully unlock the potential of THz networks for ultra-high data rates.
\bibliographystyle{IEEETran}
\bibliography{IEEEabrv,references}

\end{document}